%% file: main_arxiv.tex
\documentclass[sigconf]{acmart}
\renewcommand\footnotetextcopyrightpermission[1]{} 
\usepackage[english]{babel}
\usepackage{amsmath}
\usepackage{float}
\allowdisplaybreaks
\usepackage{enumitem}
\restylefloat{table}
\usepackage{float}

\usepackage{amsmath}
\usepackage{graphicx}
\usepackage{indentfirst}

\AtBeginDocument{%
  }



\title[BEAR: Physics-Principled Building Environment for Control and RL]{BEAR: Physics-Principled Building Environment for Control and Reinforcement Learning }

\author{Chi Zhang, Yuanyuan Shi}
\affiliation{ \department{Department of Electrical and Computer Engineering}
\institution{University of California San Diego}}
\author{Yize Chen}
\affiliation{\department{AI Thrust, Information Hub}
\institution{Hong Kong University of Science and Technology}
}

\begin{document}
\begin{abstract}
Recent advancements in reinforcement learning algorithms have opened doors for researchers to operate and optimize building energy management systems autonomously. 
However, the lack of an easily configurable building dynamical model and energy management task simulation and evaluation platform
has arguably slowed the progress in developing advanced and dedicated reinforcement learning (RL) and control algorithms for building operation tasks. Here we propose “BEAR”, a physics-principled Building Environment for Control and Reinforcement Learning. The platform allows researchers to benchmark both model-based and model-free controllers using a broad collection of standard building models in Python without co-simulation using external building simulators. In this paper, we discuss the design of this platform and compare it with other existing building simulation frameworks. We demonstrate the compatibility and performance of BEAR with different controllers,  including both model predictive control (MPC) and several state-of-the-art RL methods with two case studies. BEAR is available at \url{https://github.com/chz056/BEAR}. 
\end{abstract}
\maketitle

\section{Introduction}
\input{intro}

\section{Building Dynamics}
\label{sec:building_dyn}
\input{dynamics}

\section{RL Environment Design}
BEAR enjoys flexibility and high fidelity provided by a variety of user-defined variables as inputs~(see details in Appendix C) and provides an OpenAI Gym interface. Users can perform simulations in the customized environment with any classic model-based control or learning-based controllers. A sample usage of BEAR package in Python is also illustrated in Appendix D.

\textbf{State Space}: The state $s_k$ is the RL agent's observation from a building environment at timestep $k$. It is different from the state space model \eqref{eq:discrete_building_model} by including both $x[k]$ and the uncontrollable inputs observed from the environment. The state space is bounded by user-defined minimum and maximum values. The state $s_k$ is constructed as:
\begin{equation}s_k=[T_1[k],T_2[k], ...,T_{M}[k],Q_p[k],T_G[k],T_{E}[k],Q_{ghi}[k]],
\end{equation}

\textbf{Action space}: 
The action $a_k$ is generated by the controller given state $s_k$. 
The action $a_k$ is a set of controllable actions constructed with the energy supply of the HVAC system, as shown below:
\begin{equation}
a_k=[Q^z_{1}[k],Q^z_{2}[k],...,Q^z_{M}[k]],
\end{equation}
The whole action space is constrained by the maximum HVAC power consumption and normalized within the box of [-1, 1] in the user interface. All actions are rescaled to their original values inside the BEAR simulation. Once an action is selected, the $env.step()$ function provided by OpenAI Gym will take both $s_k$ and $a_k$ as input, and simulate the next state $s_{k+1}$ using BEAR. 

\begin{figure}[t]
\centering
\includegraphics[width=0.5\textwidth]{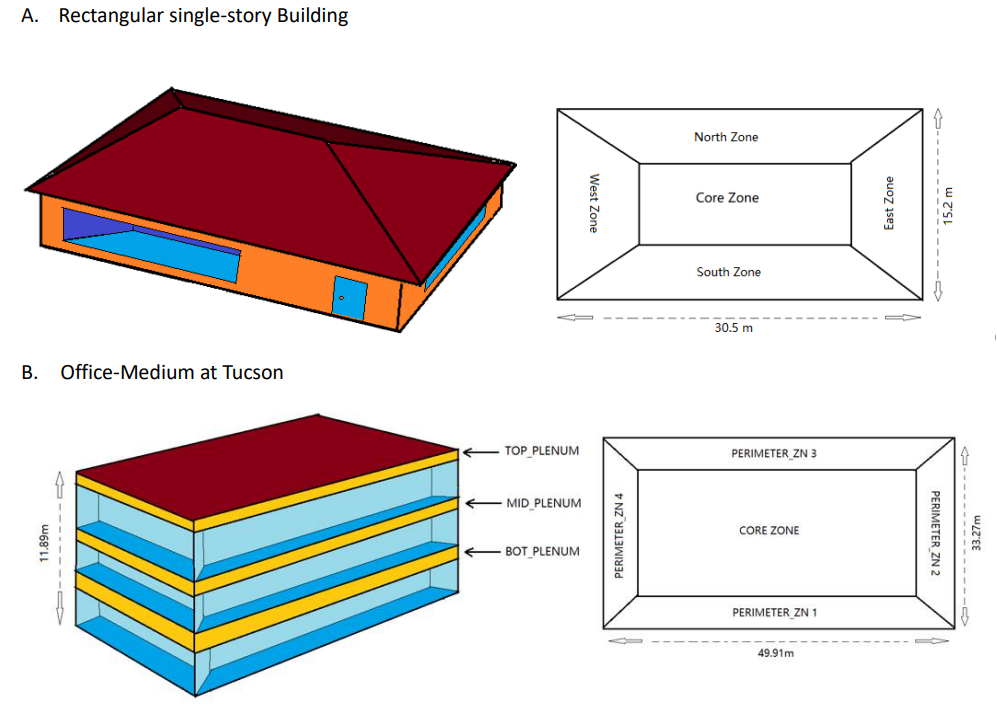}
\caption{\label{fig:rp18b} Simulated building examples.\vspace{-10pt}}
\end{figure}

\textbf{Reward}:
A main objective of building control is to reduce the energy consumption while keeping the temperature within given comfort range. 
Our platform allows users to customize reward function using environment states $s_k$, actions $a_k$, target values $obj_k$ and a weight term $\beta$. We denote such reward function as 
$Reward = R(s_k,a_k,obj_k,\beta)$.
BEAR users can customize reward by changing the weighting term, with small $\beta$ leading to low energy consumption and large $\beta$ leading to small temperature range deviation. One default reward function is the \emph{L2} reward, defined as $$R_k=-(1-\beta)||a_k||_2-\beta||s^{obj}_k-s_k||_2$$ where $s^{obj}_k=[T^{obj}_{1}[k],...,T^{obj}_{M}[k]]$ are the target temperature from user input.

\section{Experiments}
\label{sec:simulation}
In this section, we demonstrate the usage of BEAR with two building examples. We compare our simulator with EnergyPlus on a rectangular single-story building to validate the fidelity of our simulations. We also compare the performance of different control strategies, including rule-based controller, MPC (which knows the exact building dynamics model), and two RL controllers SAC~\cite{haarnoja2018soft}, PPO~\cite{schulman2017proximal} (which do not assume any model knowledge) on a simulated medium office building. Since our goal is to show the compatibility of our platform with multiple controllers, we directly set up all RL algorithms using Stable-Baselines 3~\cite{stb3}. Detailed variable settings and building type examples can be found in our anonymous code repository.

\begin{figure}[t]
\centering
\includegraphics[width=0.5\textwidth]{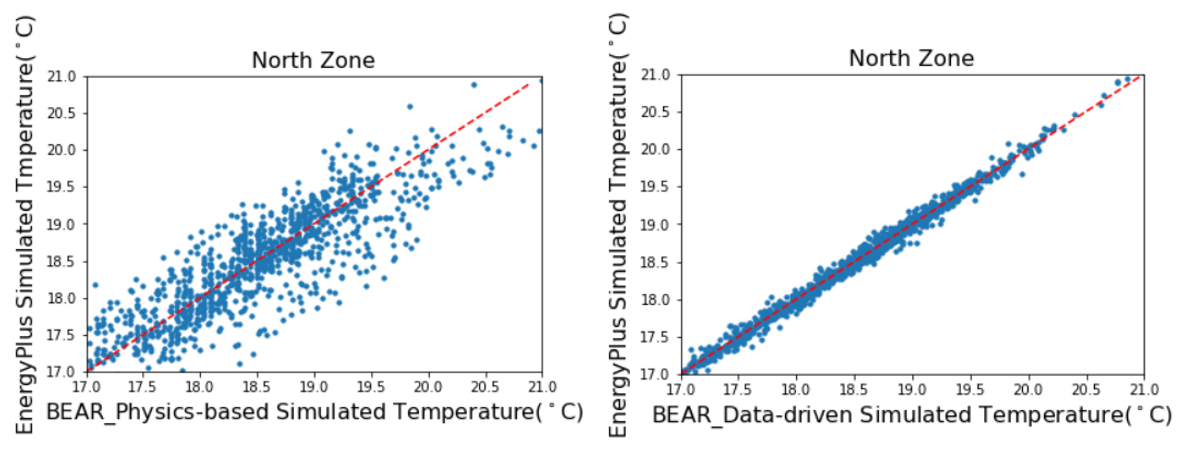}
\caption{\label{fig:SvA}Comparison of zonal temperature using BEAR simulator and EnergyPlus on the same test building located in Chicago, IL.\vspace{-10pt}}
\end{figure}

\begin{figure}
\centering
\includegraphics[width=0.5\textwidth]{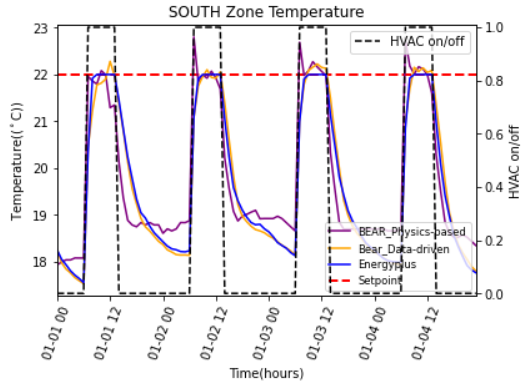}
\caption{\label{fig:south} Temperature profile comparison with same control.\vspace{-10pt}}
\end{figure}

\begin{figure}
\centering
\includegraphics[width=0.5\textwidth]{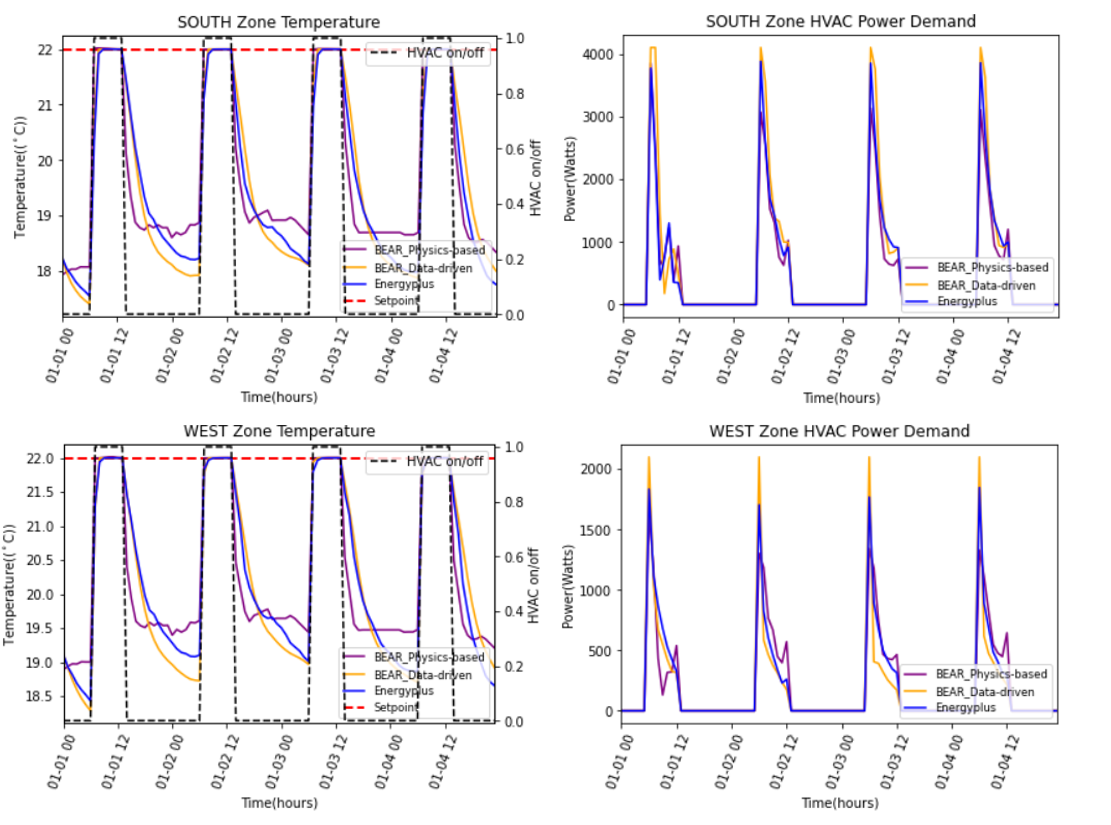}
\caption{\label{fig:paper2} Comparison of energy consumption. \vspace{-10pt} }
\end{figure}

\textbf{Single-Story Building}:  
To illustrate the fidelity of our simulator, we first set up a test example building model described in the EnergyPlus documentation and benchmark it against BEAR's simulations. 
See Fig \ref{fig:rp18b} (A) for the building illustration. Specifically, we set up the physics-based model with building parameters estimated through information provided in the EnergyPlus. We also fit a data-driven model with data simulated by the EnergyPlus. A comparison of the zonal temperature simulated in BEAR and EnergyPlus without any HVAC control is shown in Fig \ref{fig:SvA}. We could see both physics-based and data-driven simulation engine demonstrate good fidelity compared to EnergyPlus. Then, we set the indoor temperature at $22^\circ$C with a daily operating schedule of 8 a.m. to 3 p.m. In Fig \ref{fig:south}, we compare the south zone temperature between our simulator and EnergyPlus using the same control input actions. We also compare the power demand by controlling each zone in the building strictly at the same target temperature. As is shown in Fig \ref{fig:paper2}, under the same operational goal, we validate that not only the daily temperature in both simulations exhibit the same patterns (left), but the energy consumption profile of BEAR also closely tracks EnergyPlus's simulated trajectory (right).

\begin{scriptsize}
\begin{table}[t]
\centering
\begin{tabular}{l||p{1.7cm}|p{1.7cm}|p{1.3cm}}
\hline
Controller & Average Temperature Variation($^\circ$C) & Average Daily Energy Consumption (J)& Computation Time (s)\\\hline
Rule-Based Controller&2.537  &1.490E6&0.698 \\\hline

MPC ($\beta=0.8$)&2.701E-11 &6.25E5 &33.572\\\hline
PPO ($\beta=0.8$) &0.969 & 6.504E5&1.309  \\\hline
SAC ($\beta=0.8$) &0.795 & 6.188E5&1.348 \\\hline
MPC ($\beta=0.45$)&0.383 &5.87E5 &33.633\\\hline
PPO ($\beta=0.45$) &2.645 & 4.680E5&1.339 \\\hline
SAC ($\beta=0.45$) &1.360 & 5.691E5&1.235 \\\hline
\end{tabular}
\caption{\label{tab:perform}Performance on medium office. \vspace{-15pt}}
\end{table}
\end{scriptsize}

\textbf{Medium Office Building}: 
We also test a medium office building provided by the reference commercial buildings list of the U.S. Department of Energy~\cite{deru2011us}. The building has three stories; each is divided into four perimeter zones, one core zone, and one plenum zone, see Fig \ref{fig:rp18b} (B). 
The HVAC system is operated at all perimeter zones, and we set the indoor temperature of the perimeter zones at $22^\circ$C. We set the location of the building in Tucson, Arizona, and use the weather file of 2003 from January to April for simulation. 
The control agents we tested include a rule-based controller, a MPC, and two RL controllers, namely PPO and SAC. The rule-based controller performs heating when the indoor temperature is below the target setpoint and performs cooling when the temperature is above the setpoint. The MPC, PPO, and SAC controllers are tested with two $L_2$ reward functions with $\beta=0.8$ and $\beta=0.45$ respectively. 
The learning curves of the RL controllers are shown in Appendix A in Fig \ref{fig:trainreward}. We could see that all the RL training rewards are converging, and algorithms with the same reward function converge to a similar objective value. The performance of tested controllers is shown in Table~\ref{tab:perform}. With the implementation of RL algorithms, we can observe that the energy consumption decreases significantly compared to the simple rule-based controller. It also reduces the average temperature variation. Compared to the MPC method, which has complete model knowledge, both model-free RL algorithms obtain a similar or lower energy consumption, with higher temperature violation. The computation time using PPO and SAC is greatly reduced compared to MPC, as the latter needs to solve an optimization problem at each step to obtain action.


\section{Conclusion}
This paper presents BEAR, an open-source physics-principled building control and RL platform compatible with OpenAI Gym. Unlike many existing platforms that use co-simulation with outsourcing building simulators, our platform is self-contained and thus provides simplicity for learning algorithm development and customized tasks. We illustrate the flexibility and efficiency of BEAR and various usage under both physics-based and data-driven settings.
We plan to fill the gap between our building model and the real-world buildings by addressing factors such as shadow and light. We also plan to extend BEAR by supporting multi-agent RL training and heterogeneous reward design. 

\bibliography{reference}
\bibliographystyle{ACM-Reference-Format}
\pagebreak[2]
\appendix

\input{appendix}

\end{document}

%% file: intro.tex
Building is one of the major sources of global energy consumption. In 2021, residential and commercial buildings were responsible for around 39\% of total U.S. energy consumption and 74\% of total U.S. electricity consumption~\cite{nalley2021annual}. Consequently, research on the operation of building Heating Ventilation and Air Conditioning (HVAC) systems can lead to significant energy savings and carbon emission reduction.
Many control methods have been developed to provide solutions for building HVAC control problems, including model predictive control, nonlinear adaptive control, and decentralized control~\cite{ma2012predictive}\cite{tang2019model}.
However, most such approaches require detailed and exact building dynamics models, and an increase in the complexity of building dynamics would lead to significantly higher computational costs. As a result, reinforcement learning (RL) has gained tremendous interest for building control in modern days due to its model-free nature (see~\cite{mason2019review} for a recent review). 

One challenge of the building RL research is the lack of a bench-marking simulation environment for developing and evaluating different RL algorithms with realistic building models. Several recent works~\cite{jimenez2021sinergym, scharnhorst2021energym,zhang2018practical,arroyo2021openai,vazquez2020citylearn} have proposed simulation solutions to address such a problem.
However, most of them adopted a co-simulation framework with a python interface for algorithm development and an outsourcing building simulator,
like EnergyPlus~\cite{crawley2000energy} or Modelica~\cite{mattsson1997modelica}. 
For researchers who do not yet have detailed knowledge of such packages, it is hard to test with their own configurations and validate RL performance. 
BOPTESTS-Gym~\cite{arroyo2021openai}, Sinergym~\cite{jimenez2021sinergym}, and Gym-Eplus~\cite{zhang2018practical} rely on EnergyPlus or Modelica to perform simulation, and the BCVTB middleware~\cite{wetter2008building} to communicate between simulators and the platform interface. Energym~\cite{scharnhorst2021energym} uses predefined building models and co-simulation with EnergyPlus. 
CityLearn~\cite{vazquez2020citylearn} is almost self-contained that it uses pre-simulated data. However, CityLearn focuses on building-level control interacting with the grid,
rather than zone-level detailed building simulation. 

In this paper, we present BEAR, a physics-principled \underline{B}uilding  \underline{E}nviron-ment for control \underline{A}nd \underline{R}einforcement learning.  BEAR constructs building simulation from first-principled physics models and 
provides a scalable platform for researchers from different backgrounds to design, test and evaluate their reinforcement learning (RL) and control algorithms. BEAR can set up customized building environments by either choosing from a curated list of (building type, weather type, and city), or incorporating building and weather datasets of their own. To address the potential gap between the simulated configurations and the actual building dynamics, BEAR users can also efficiently train a data-driven model using self-collected dataset of their own. The proposed simulator supports fine-grained dynamics simulation and provides an OpenAI Gym interface~\cite{brockman2016openai} for developing RL agorithms. Researchers from the machine learning and RL community can design new environments and algorithms with minimal knowledge of the underlying building physical models and thus can focus more on algorithm development and evaluation. 
On the other hand, BEAR, with a physics-principled simulation engine, provides researchers and engineers an accessible platform to implement new building models with user-defined building structure, operation schedule, temperature, and other environmental variables. The primary characteristic that distinguishes BEAR from other building RL simulators is the physics-based modeling procedures as described in Section~\ref{sec:building_dyn}.
Table \ref{tab:lala} compares BEAR with some other RL environments for building control. \vspace{-10pt}


\begin{scriptsize}
\begin{table*}
\centering
\begin{tabular}{p{2cm}|p{1.5cm}|p{1.5cm}|p{1.5cm}|p{1.5cm}|p{1.5cm}|p{2cm}|p{2cm}}\hline
    & BEAR & Sinergym~\cite{jimenez2021sinergym}& Energym~\cite{scharnhorst2021energym} & Gym-Eplus~\cite{zhang2018practical} & Citylearn~\cite{vazquez2020citylearn}&RL Testbed for
EnergyPlus~\cite{moriyama2018reinforcement}&BOPTESTS-Gym~\cite{arroyo2021openai}
 \\\hline\hline
Simulator & Self-designed & EnergyPlus & EnergyPlus & EnergyPlus & Data & EnergyPlus& Modelica\\\hline
Available Buildings & 16+ & 3 & 7 & 1 & 9&1&6\\\hline
Weather Types & 19 & 3 & 4 & 1 & 5&5&5 \\\hline
Action Space & Both & Both & Discrete & Discrete & Continuous &Continuous&Both\\\hline
Action Type & Energy & Temperature & Temperature & Temperature & Energy&Temperature \& Fan flow rate& Temperature \& Lower level actuator signals\\\hline
Reward & Customized & Customized & Customized & Customized & Predefined& Predefined&Customized\\\hline
Multi-agent & User-defined & No & No & No & Yes& No&No\\\hline
Control Objectives & Energy demand, Thermal comfort & Energy demand, Thermal comfort & Grid exchange, Energy demand, CO2 emissions & Energy demand, Thermal comfort & Energy demand&Energy demand, Thermal comfort&Energy demand, Thermal comfort\\\hline
Control Step & User-defined & User-defined & User-defined & 5 minutes & 1 hour & 15 minutes& User-defined\\\hline
Zone level Control & Yes & No & Yes & No & No&Yes&Yes\\\hline
\end{tabular}
\caption{\label{tab:lala} Features of building control simulators.}
\end{table*}
\end{scriptsize}


%% file: dynamics.tex
The Reduced Resistance-Capacitance (RC) model~\cite{ma2012predictive} is widely used for the building HVAC system model to simplify design complexity and reduce computation time. We construct the BEAR physics-based building simulation model based on the RC model with an nonlinear residual model. 
BEAR enables three user inputs: Building type, Weather type, and City. Users can either choose from a pre-defined list of buildings and climate types provided by Building Energy Codes Program~\cite{buildingenergy} (See Appendix B) or define a customized BEAR environment by importing any self-defined EnergyPlus building models and weather files. BEAR also explicitly incorporates the nonlinear and stochastic heat transfer caused by building occupancy, making it flexible in considering various building usage and control scenarios.

\begin{figure}[t]
\centering
\includegraphics[width=0.37\textwidth]{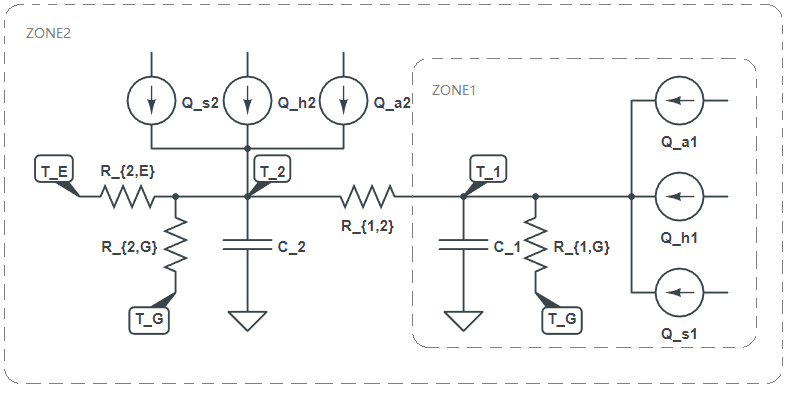}
\caption{\label{fig:RCModel}RC Model for a two-zone, one-story building.\vspace{-15pt}}
\end{figure}
We illustrate our simulator design using a two-zone, one-story building model as an example, as shown in Fig \ref{fig:RCModel}. ZONE 1 is entirely inside ZONE 2; thus, only ZONE 2 has external walls that connect to the outside air. The heat transferred through the model is considered from the temperature difference between neighboring zones, occupants' activity ($Q^a$), global horizontal irradiance ($Q^s$), and HVAC systems ($Q^h$). 
Here we show the modeling process for ZONE 1 and ZONE 2 using the following differential equations based on the RC model~\cite{ma2012predictive}:
\begin{subequations}\label{eq:zone_temp_ode}
    \begin{align}
    C_{1}\frac{dT_1}{dt} 
    &=\frac{T_2-T_1}{R_{1,2}}+\frac{T_G-T_1}{R_{1,G}}+Q^h_{1}+Q^a_{1}+Q^s_{1};\\
    C_{2}\frac{dT_2}{dt} 
    &=\frac{T_{E}-T_2}{R_{E,2}}+\frac{T_1-T_2}{R_{1,2}}+\frac{T_G-T_2}{R_{2,G}}+Q^h_{2}+Q^a_{2}+Q^s_{2},
    \end{align}
\end{subequations} 
where $T_i$ is Zone $i$'s temperature, $T_G$ and $T_E$ denote the ground and outdoor environment temperature.
$C_i$ is the thermal capacitance, $R_{i,j}$ is the thermal resistance between Zone $i$ and Zone $j$ and is symmetric, i.e., $R_{i,j}=R_{j,i}$. 
$Q^{h}_i$ is the controlled heating supplied to each zone;
$Q^{a}_i$ is the heat gained from indoor people activities; 
$Q^{s}_i$ is the solar heat gained from windows for Zone $i$.

For a general building model with $M$ indoor zones $i = 1, 2, ..., M$, the zone thermal dynamics are as follows:
\begin{equation}
    \begin{aligned}
    C_{i}\frac{dT_i}{dt} 
    =\sum_{j \in \mathcal{N}(i)} \frac{T_{j}-T_{i}}{R_{i, j}}+Q^h_{i}+Q^a_{i}+Q^s_{i},
    \end{aligned}
\end{equation}
where $\mathcal{N}(i)$ are the neighboring zones of zone $i$.
We encode each zone's connectivity to ensure only ground floor zones are connected to Zone $G$ (ground), and peripheral zones are connected to Zone $E$ (outdoor environment).

\textbf{Heat Transfer Modeling}:  The following equations are descriptions of heat gained from different sources:
\begin{equation}
    \quad Q^{h}_i=w_{i}Q^z_{i}; \quad Q^{a}_i = n_{i}Q_{p}; \quad  Q^{s}_i = \alpha_{SHGC} A^{win}_i Q_{ghi},
\end{equation}
 where $w_i$ represents the zonal HVAC efficiency coefficient, and $Q^z_{i}$ denotes the heat gained from HVAC to compute the controlled zone heating. For modeling the heat gain from human activities $Q^{a}_i$, $n_i$ denotes the number of people in each zone, and $Q_p$ is the sensible heat gained from activities by one person. To model the solar heat, $\alpha_{SHGC}$ refers to the solar heat gain coefficient for windows, $A^{win}_i$ is the zonal window area, and $Q_{ghi}$ is the heat absorbed from global horizontal irradiance. Both $Q^{a}_i$ and $Q^{s}_i$ are time-varying uncontrolled heat generated from the environment, while $Q^{h}_i$ is the controllable heat that could be taken as inputs from the simulator.  

Here we calculate sensible heat per person $Q_p$ using a polynomial function detailed in the EnergyPlus documentation~\cite{epef}:
\begin{equation}\label{eq:Q_p}
    \begin{aligned}
    Q_p= & c_1+c_2m+c_3m^2+c_4\bar{T}-c_5\bar{T}m+c_6\bar{T}m^2\\
    &-c_7\bar{T}^2+c_8\bar{T}^2m-c_9\bar{T}^2m^2,
    \end{aligned}
\end{equation}
where $m$ is the metabolic rate, $\bar{T} = \frac{1}{M} (T_1 + T_2 + ... + T_M)$ is the average zone temperature, and $c_1, ..., c_9$ are constants generated by fitting sensible heat data under varying conditions.

\textbf{State Evolution:} to simulate the building with designed control inputs, we re-organize the system dynamics model in \eqref{eq:zone_temp_ode}-\eqref{eq:Q_p} to the state-space form:
\begin{equation}\label{eq:building_statespace}
    \begin{aligned}
    \dot{{x}}(t) &= {A} {x}(t) +{B} {u}(t) + {D}f(t,x(t),r(t)),
    \end{aligned}
\end{equation}
where the state variable represents the collection of zone temperature,
$x(t)=[T_1  \; T_2  \;\cdots \; T_M]^\top$, the control variable $u(t)=[T_G \; T_{E} \; Q^z_{1} \; Q^z_{2} \; \cdots \; Q^z_{M}\; Q_{ghi}]^\top$, and the nonlinear function for sensible heat calculation 
$f(t,x(t),r(t))=c_1+c_2r(t)+c_3r(t)^2-c_5\bar{T}r(t)+c_6\bar{T}r(t)^2-c_7\bar{T}^2+c_8\bar{T}^2r(t)-c_9\bar{T}^2r(t)^2 $, where $r(t)$ is the current metabolic rate $m$. 
The state matrix is as follows
\begin{tiny}
$$A=
\setlength{\arraycolsep}{1pt}
\begin{bmatrix}
 \sum_{j \in \mathcal{N}(1)} \frac{-1}{C_1R_{1, j}} + \frac{c_4 n_1}{MC_1}&\frac{1}{C_1R_{1,2}} + \frac{c_4 n_1}{MC_1} &\dots&\frac{1}{C_1R_{1,M}} + \frac{c_4 n_1}{MC_1}\\
\frac{1}{C_2R_{2,1}} + \frac{c_4 n_2}{MC_2}&\sum_{j \in \mathcal{N}(2)} \frac{-1}{C_2R_{2, j}} + \frac{c_4 n_2}{MC_2} &\dots&\frac{1}{C_2R_{2,M}} + \frac{c_4 n_2}{MC_2}\\
\vdots& &\ddots& &\\
\frac{1}{C_MR_{M,1}} + \frac{c_4 n_M}{MC_M}&\dots& \dots &\sum_{j \in \mathcal{N}(M)} \frac{-1}{C_MR_{M, j}} + \frac{c_4 n_M}{MC_M}\\
\end{bmatrix},
$$
\end{tiny}
the input matrix is
\begin{tiny}
$$B=
\setlength{\arraycolsep}{3pt}
\begin{bmatrix}
 \frac{1}{C_1R_{1,G}} & \frac{1}{C_1R_{1,E}}& \frac{w_1}{C_1} &0&\dots&0&\frac{\alpha_{SHGC} A^{win}_{1}}{C_1}\\
\frac{1}{C_2R_{2,G}}&\frac{1}{C_2R_{2,E}} &0&\frac{w_2}{C_2}&\dots&0&\frac{\alpha_{SHGC} A^{win}_{2}}{C_2}\\
\vdots&\vdots&\vdots& &\ddots& &\vdots\\
\frac{1}{C_MR_{M,G}}&\frac{1}{C_MR_{M,E}} &0&0&\dots&\frac{w_M}{C_M}&\frac{\alpha_{SHGC} A^{win}_{M}}{C_M} \\
\end{bmatrix},
$$
\end{tiny}and matrix $D$ in Eq~\eqref{eq:building_statespace} is calculated as ${D}=[\frac{n_1}{C_1}  \; \frac{n_2}{C_2}  \;\cdots \; \frac{n_M}{C_M}]^\top
$. We convert the continuous-time system model into a discrete-time representation, 
\begin{equation}\label{eq:discrete_building_model}
    x[k+1] = A_dx[k]+B_du[k] + D_dF[k,x[k], r[k]],
\end{equation}
where the term $A_d = e^{A \Delta T}$, $\Delta T$ is the sample time resolution. 
$B_d = A^{-1}(A_d-I)B $, $D_d= A^{-1}(A_d-I)D$ and $F[k,x[k], r[k]] = \int_{k \Delta T}^{(k+1) \Delta T} f(\tau,x(\tau), r(\tau)) d \tau $.



BEAR enables an automated pipeline to process building geometry, weather, and occupancy information to the discrete-time state-space models. Building parameters are obtained through the user-input building information. For example, $R$ is determined by the wall material and volume, $C$ is estimated by the volume of each zone, $Q$ is computed by a combination of occupancy, GHI, and VAV information. Compared to the actual building model, our model in Eq~\eqref{eq:discrete_building_model} makes several simplifications regarding the zone shape, the window/door open schedules, the heat transfer function of HVAC, and the shadowing function. Detailed model assumptions are listed in the linked code repository. Nevertheless, extending the pre-defined dynamics to building use cases with user-defined zone shape, schedules, and shadowing functions is adaptable with our open-source building simulation engine.

\textbf{Data-driven model}: In some practical scenarios, the building parameters are not known exactly a priori, while only historical power consumption and temperature measurements are available. 
To address such gap between the simulation parameters and the actual building dynamics, we also incorporate a data-driven module in BEAR. We use Linear Regression to fit a data-driven building model $\hat{Y} =  WX + b $ with coefficients $\{W,b\}$ that minimize the residual sum of squares between the ground truth and the predicted building states. Users could train with state and action data collected from a particular building of interests with minimal efforts and use the data-driven building model for controller/RL algorithm design. Besides the default linear data-driven models, users can also define other types of data-driven building models in BEAR such as neural networks and run gradient descent to update the model estimates. 

BEAR's data-driven module takes the current step state-action pair $x[k],u[k],r[k]$ as input and predicts the next time-step state $x[k+1]$. To address the non-linear part of our model,  we include $\bar{T}^2$ from the nonlinear function $F$ as one of the input features, which could be calculated using $x[k]$. Thus, we collect $X=[x[k],u[k],\bar{T}^2]$ with target $Y=x[k+1]$.

%% file: appendix.tex
\begin{scriptsize}
\begin{table*}
\centering
\begin{tabular}{l||p{3cm}|p{5cm}|p{3cm}|p{3cm}}
Input Variables & Type & Description & Default value & Source   \\\hline\hline
Filename & Str & Filename of the selected building model & Required & Building List or User-defined \\\hline
Weatherfile & Str & Filename of the selected weather & Required & Weather list \\\hline
Location & List with length of 12  & Ground temperatures of 12 month& Required & Ground Temperature Dictionary \\\hline
U-Wall & List with length of 7 & U-factors of Walls & Given by building model & Building list \\\hline
Target &List with length of zone number &  Target temperature of comfort & [$22^\circ$C...$22^\circ$C] & User-defined \\\hline
Time Reso & Int & Length of one time-step & 3600 second & User-defined  \\\hline
Reward-gamma & List with length of 2 & Weight for comfort level and energy demand &[0.001,0.999] & User-defined \\\hline
SHGC & Int & Solar Heat Gain Coefficient & 0.252 & User-defined \\\hline
SHGC-weight &Int& Radiative/convective split for heat gain& 0.1 & User-defined  \\\hline
Ground-weight & Int&Lost of heat gain from ground & 0.5 & User-defined \\\hline
Full-Occ & List with length of zone number & Number of people in each zone & [0...0]person & User-defined \\\hline
Activity-sch& List with length of the simulation & The activity schedule of people & [120...120] W/person & User-defined\\\hline
AC-map & List of Boolean with length of zone number& Map of HVAC in the building &[1...1]&User-defined \\\hline
Max-power &Int& Maximum power of HVAC & 8000 W & User-defined \\\hline
\end{tabular}
\caption{\label{tab:widgets}Variables table.}
\end{table*}
\end{scriptsize}

\section{RL Training Curve}
\begin{figure}
\centering
\includegraphics[width=0.4\textwidth]{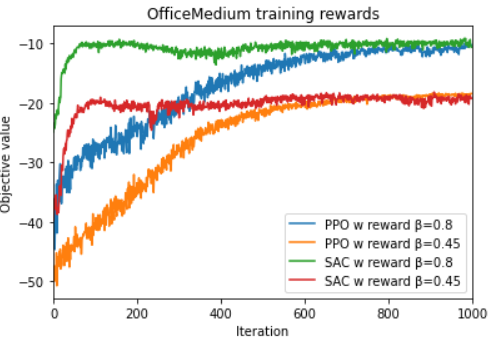}
\caption{\label{fig:trainreward}Reward during Training for RL policies.\vspace{-10pt} }
\end{figure}

Training curve for the PPO and SAC algorithms with different rewards in controlling the medium office building in figure ~\ref{fig:trainreward} .

\section{List of Simulating Configurations}
The prototype buildings included in BEAR are derived from DOE's Commercial Reference Building Models. The models include 16 commercial building types in 19 locations. Users can download the models at \url{https://www.energycodes.gov/prototype-building-models }. Since BEAR is compatible with EnergyPlus, users could also create their own model in EnergyPlus editor and load the generated table file into BEAR.

\begin{itemize}[leftmargin=*]
    \item \emph{Available building types}: ApartmentHighRise, ApartmentMidRise, Hospital, HotelLarge, HotelSmall,  OfficeLarge, OfficeMedium, OfficeSmall, OutPatientHealthCare, RestaurantFastFood, RestaurantSitDown, RetailStandalone, RetailStripmall, SchoolPrimary, SchoolSecondar, Warehouse.
    \item \emph{Available weather types}: Very Hot Humid, Hot Humid, Hot Dry, Warm Humid, Warm Dry, Warm Marine, Mixed Humid, Mixed Dry, Mixed Marine, Cool Humid, Cool Dry, Cool Marine, Cold Humid, Cold Dry, Very Cold, Subarctic/Arctic.
    \item \emph{Available locations}: Albuquerque, Atlanta, Buffalo, Denver, Dubai, ElPaso, Fairbanks, GreatFalls, HoChiMinh, Honolulu, InternationalFalls, NewDelhi, NewYork, PortAngeles, Rochester, SanDiego, Seattle, Tampa, Tucson.  
\end{itemize}


\section{Variable table}
A large variety of variables could be defined by BEAR users. Three inputs (Filename, Weatherfile, Location) are required for setting up a basic building environment, while users could modify other variables for better simulation accuracy. Variable \textbf{U-WAll} contains the U values of each wall used in the building model, which could be changed if user would like to replace the material of the wall. Variable \textbf{Target} should be used for self-define control objectives. Variable \textbf{Time-Reso} should be modified if user would like to change the sample time of the simulation. Variable \textbf{Reward-gamma} can change  the reward function coefficient. Variable \textbf{SHGC} is based on the materials of window. Both SHGC and Ground temperature could have partial impact on zone temperature depending on the building structure, thus \textbf{SHGC-weight} and \textbf{Ground-weight} are provided for user to tune. Variable \textbf{Full-Occ} is used to set up occupancy of each zone. Variable \textbf{Activity-sch} represents the metabolic heat of different activities. Variable \textbf{AC-map} and \textbf{Max-power} can be used to change the location and the maximum output of the HVAC system. To address the non-linear part of our model, we assume all people in the building perform similar activities, which guarantees a constant metabolic rate $r(t)$.

Table~\ref{tab:widgets} summarizes the full list of variables currently implemented in BEAR.

\section{Example usage}
\begin{figure}[htbp]
\centering
\includegraphics[width=0.5\textwidth]{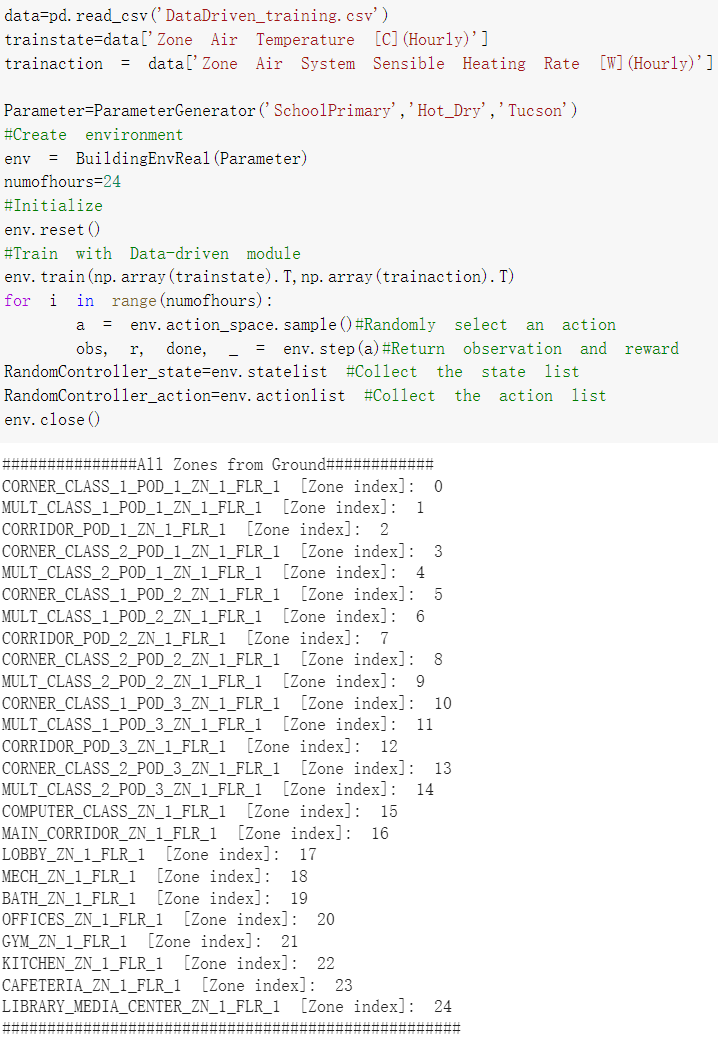}
\caption{\label{fig:code}Code for a sample usage of BEAR.}
\end{figure}
\begin{figure}[htbp]
\centering
\includegraphics[width=0.5\textwidth]{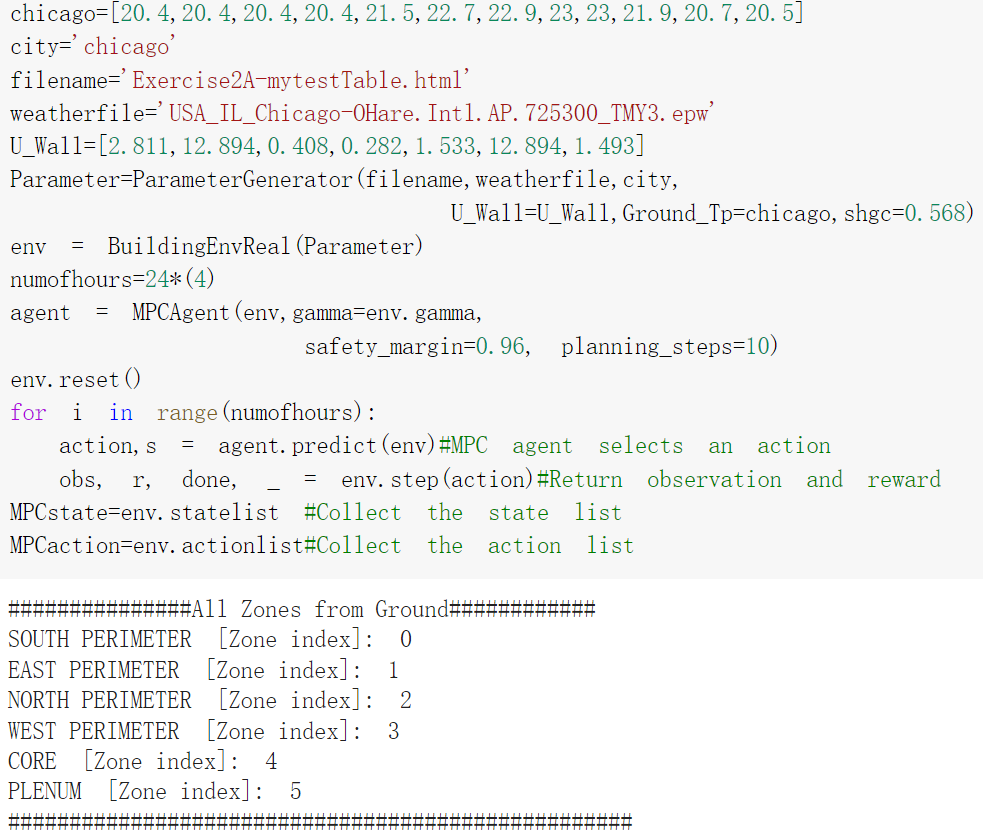}
\caption{\label{fig:code2}Code for a customize usage of BEAR.}
\end{figure}
A simple usage example is shown in Fig \ref{fig:code}. The objective of the code snippet is to simulate a 'SchoolPrimary' type building at Tucson with 'Hot Dry' weather using random selected actions. As a first step, an environment with the required building/weather/city is created with the building parameters generated. Once a building model is created, the detailed Zone information would be printed out. Then we start the simulation with the reset function to observe the initial state. 
At each step, the controller agent would observe the current state $s_t$, and generate a corresponding action $a_t$. The environment would then take the action and pass it into the building state-space model to simulate the new state $s_{t+1}$ for the next timestep. A 24-hour simulation is performed in the for-loop in this example case. Each loop would generate a random action, and send the action to the environment to observe the new state, reward, and termination.


A customized usage example for self-defined building is shown in Fig \ref{fig:code2} to illustrate. In this code snippet, a new building not included in the provided building prototype list is shown. User can self-define a building through the EnergyPlus editor, and upload the EnergyPlus html file and the epw weather file into BEAR. In the example, the wall materials and ground temperatures are customized. SHGC value is also modified. During simulation, a MPC controller is used instead of a RL controller.